\documentclass[conference]{IEEEtran}
\IEEEoverridecommandlockouts
\usepackage{cite}
\usepackage{amsmath,amssymb,amsfonts}
\usepackage{algorithmic}
\usepackage{graphicx}
\usepackage{textcomp}
\usepackage{xcolor}
\usepackage{hyperref}
\usepackage{siunitx}
\usepackage{float}
\usepackage{fixltx2e}
\usepackage{tabularx}
\usepackage{listings}
\usepackage{ragged2e}
\usepackage{arydshln}
\usepackage{indentfirst}

\def\BibTeX{{\rm B\kern-.05em{\sc i\kern-.025em b}\kern-.08em
    T\kern-.1667em\lower.7ex\hbox{E}\kern-.125emX}}
\begin{document}


\title{A domain-specific modeling and analysis environment for complex IoT applications 
\thanks{This work has received funding from the Lowcomote project under European Union’s Horizon 2020 research and innovation program under the Marie Skłodowska-Curie grant agreement n\si{\degree} 813884.}
}

\author{

\IEEEauthorblockN{1\textsuperscript{st} Felicien Ihirwe}
\IEEEauthorblockA{\textit{Innovation Technology Services} \\
\textit{Intecs Solutions Spa}\\
Pisa, Italy\\
felicien.ihirwe@intecs.it}

\and

\IEEEauthorblockN{2\textsuperscript{nd} Davide Di Ruscio}
\IEEEauthorblockA{\textit{DISIM } \\
\textit{University of L'Aquila}\\
L'Aquila, Italy\\
davide.diruscio@univaq.it}

\and
\IEEEauthorblockN{3\textsuperscript{rd} Silvia Mazzini}
\IEEEauthorblockA{\textit{Innovation Technology Services} \\
\textit{Intecs Solutions Spa}\\
Pisa, Italy\\
silvia.mazzini@intecs.it}

\and

\IEEEauthorblockN{4\textsuperscript{th} Alfonso Pierantonio}
\IEEEauthorblockA{\textit{DISIM } \\
\textit{University of L'Aquila}\\
L'Aquila, Italy\\
alfonso.pierantonio@univaq.it}
}

\maketitle
\vspace{-.2cm}
\begin{abstract}
To cope with the complexities found in the Internet of Things domain, designers and developers of IoT applications demand practical tools. 
Several model-driven engineering methodologies and tools have been developed to address such difficulties, but few of them address the analysis aspects. In this extended abstract, we introduce CHESSIoT, a domain-specific modeling environment for complex IoT applications. In addition, the existing supported real-time analysis mechanism, as well as a proposed code generation approach, are presented.
 \vspace{-.2cm}
\end{abstract}

\section{Introduction}
Designing and developing  IoT applications require efficient tools to deal with the intrinsic complexities of this kind of systems. The current trend of domain-specific development platforms unravels the opportunities to advance simplicity on how new applications are developed in different application domains, including database systems, mobile applications, process applications, and request-handling applications. IoT is expected to become an additional domain, which can be approached through low-code methods \cite{lowcomote}. However, IoT systems tend to be very large and complex in industrial settings due to the large amount of data collected and processed concurrently. Even though we see a convenient push toward coping with the design complexity, there is still a massive gap regarding the analysis, validation, and certification of such systems. 

Even though there are some model-driven engineering approaches and tools to help with the development of IoT systems, few of them provide developers with analysis features \cite{ihirwe2021towards}. In this paper, we present the CHESSIoT environment, a domain-specific modeling environment for complex IoT applications. CHESSIoT extension aims to provide a robust domain-specific language for designing and analyzing large and complex IoT systems that can be applied to use cases such as Smart Cities. The proposed environment extends the CHESS tool \cite{CHESScomplex}, a mature open-source modeling and analysis environment for complex dependable systems. The CHESS tool-set allows the modeler (user) to automate various development phases, such as requirements definition, architectural modeling of the system's software and hardware, and deployment to a hardware platform. 

Despite the fact that CHESS has been successfully used in a variety of applications \cite{CHESScomplex}, it still lacks formal modeling and analytic capabilities for the IoT domain. CHESSIoT extends and relies on the existing modeling and analysis infrastructure provided by CHESS. CHESSIoT is being developed in the context of the European Lowcomote project \footnote{\url{https://www.lowcomote.eu/}} that aims at training the next generation of experts in low-code engineering applications (LCEPs) \cite{lowcomote}. This will be achieved by injecting in low-code development platforms (LCDPs), the theoretical and technical framework defined by recent research in Model-Driven Engineering, augmented with Cloud Computing and Machine Learning techniques. The development of CHESSIoT is still in its early phases but with very significant progress concerning its design capabilities.
 \vspace{-.1cm}
\section{The CHESSIoT modeling and analysis approach}
 \vspace{-.1cm}
CHESSIoT adopts a component-based approach, in which the user decouples various IoT functional portions of the system into components that can be referred to as smart components. These components can be modeled, analyzed, verified, saved, reused, and combined to satisfy the system's shared goals. To guaranty a fully decoupled extension, CHESSIoT introduced the \textit{"IoT sub-view"}, and once applied in all design stages, at all design views, the user will benefit from a dedicated IoT-specific modeling infrastructure consisting of specific diagrams and palettes. Furthermore, the CHESSIoT modeling profile is organized in terms of different UML profiles depending on the current design stage. 

The \textit{CHESSIoTSystem profile} extends SysML, and it is used to define the design characteristics of the system under development, such as the principal blocks and their interconnections. The IoT high-level blocks and their accompanying flow-ports are defined and later annotated with formal properties as CHESS contracts. In this regard, the user can benefit from the CHESS tool's system-level validation and verification, parameterized architecture, and trade-off analysis infrastructure \cite{CHESScomplex}. The CHESSIoTSoftware profile gives users modeling constructs to define IoT software components and behaviors. In this case, the user can employ specific palettes to break down the system's software components and sub-components. State machines are used to define the component behaviors in CHESSIoT. Extra-functional properties are specified to annotate the IoT software elements for analysis purposes. 

To address the complexity in modeling the behavior aspects of the components, CHESSIoT proposes a more flexible event-based modeling mechanism. Using the standard Papyrus infrastructure, the UML state machine events and actions modeling process is normally complex and difficult for a UML beginner. \textit{IoTEvents} and \textit{IoTActions} can be modeled individually in CHESSIoT and then activated as many times as possible or by a variety of \textit{IoTElement} states. An \textit{IoTState} should always be associated with \textit{"OnEntry"} and \textit{"OnExit"} events, which can be either general, incoming, or outgoing. Sending and receiving payloads are handled by the \textit{IoTAction} element, and action types can be assigned to each event in the form of an effect. Any form of messages transmitted between \textit{IoTElements} via ports is referred to as payload, and it can be reused as many times as necessary.

The physical representation of the virtual components is performed in \textit{CHESSIoTHardware profile}. The target platform specifications, such as the number of processors and core units, are also defined as part of the hardware design. The majority of the features in this profile are dependent on UML MARTE. Finally, all of the information about the system's communication elements is contained in the \textit{CHESSIoTOperational profile}. This profile should ideally be used to model communication modes, servers, communication protocols, and storage resources. In addition, this profile will be included to aid in the performance analysis of the resource blocks in question. 

The current CHESSIoT analysis environment infrastructure relies on the exiting analysis infrastructure provided by CHESS. CHESS provides a variety of analyses depending on the needs (functional, timing, dependability). Although the current CHESSIoT models can fully comply with such analysis, we understand that the analysis mechanism might not fully align with the IoT domain settings. For instance, CHESS's real-time analysis infrastructure relies on the MAST tool \cite{MAST}, a timing analysis tool that is based on the component timing requirements of the system. The user must annotate the real-time temporal logic attributes on each component's operations to execute such analysis. The timing request type (i.e., periodic or sporadic), the request's worst-case execution time, the priority, and the desired execution deadline are among the attributes. 

It is worth noting that in IoT domain settings, the activities are more dynamic and unpredictable. This could be due to a variety of unexpected environmental constraints. We must also remember that deadlocks events might occur in the middle of a process, such as when a processing resource fails. CHESSIoT aims to build on top of the CHESS real-time analysis environment to incorporate such IoT-specific properties in the analysis processes.
\vspace{-.2cm}

\section{The code generation methodology}
The CHESSIoT environment will also offer a possibility to generate the platform-specific code. This is achieved by employing an intermediate ThingML model and later get transformed into platform-specific code. ThingML is amongst the most popular domain-specific model-driven engineering tools for the Internet of Things domain. It comprises a custom textual modeling language, a supporting modeling tool, and advanced code generator capabilities. ThingML code generator targets many popular programming languages such as C/C++, Java, and Javascript, and about ten different target platforms (ranging from tiny 8bit microcontrollers to servers) and ten different communication protocols \cite{ThingMLcore}. 

The CHESSIoT component's semantics differs from the ThingML's to some extent, and that is why the mapping of the elements is needed to solicit an efficient transformation. Table \ref{tab:extension} summarizes the mapping between the CHESSIoT model and ThingML, which is therefore referred to when performing CHESSIoT2ThingML model transformation. Even though the ThingML framework is well-developed and promising, not all aspects have been covered. For example, the ThingML framework cannot conduct any system-related analysis, which is crucial in a sector like IoT where safety is a concern. CHESSIoT proposes a centralized and fully automated environment where users may design, develop, analyze and generate IoT systems using ThingML and CHESS technologies. Note that a CHESSIoT user will not need ThingML knowledge to generate code. For full access to the CHESSIoT approach, please consult our previous work \cite{ihirwe2021towards}. \vspace{-.4cm}
\begin{table}[h]
  \caption{Proposed CHESSIoT2ThingML mapping}
  \label{tab:extension}
  \begin{tabular}{c|cl}
    \textbf{\texttt{CHESSIoT element }} && \textbf{\texttt{ThingML element}}\\\hline
    \texttt{Component} && \texttt{Thing}\\
    \texttt{Provided/required port} &&  \texttt{Provided/required port}\\
    \texttt{Operation} && \texttt{Function}\\
    \texttt{Property} && \texttt{Property}\\
    \texttt{Payload} && \texttt{Message}\\
    \texttt{IoTState/Transition} && \texttt{State/Transition}\\
    \texttt{StateGuards} && \texttt{Guards}\\
    \texttt{IoTEvent/Action} && \texttt{Event/Action}\\
  \end{tabular}
   \vspace{-.5cm}
\end{table}

\section{Conclusion}
Because of the inherent complexity of IoT systems, it is necessary to conceive techniques and tools to develop and manage them. Furthermore, early analysis of such systems can detect possible issues before system deployment and execution. This paper presents CHESSIoT, a domain-specific modeling and analysis environment for complex IoT applications. We have also introduced the methodology of the envisioned code generation approach.    
\vspace{-.2cm}
\bibliographystyle{IEEEtran}
\bibliography{IEEEabrv,Bibliography}
\vspace{-.2cm}
\end{document}